\documentclass{article} 
\usepackage{latexsym} 
\usepackage{graphicx} 
\usepackage{epsfig} 
\newcommand \be {\begin{equation}} 
\newcommand \bea {\begin{eqnarray}} 
\newcommand \ee {\end{equation}} 
\newcommand \eea {\end{eqnarray}}

\begin{document} 
\topskip 2cm 
\begin{titlepage} 
\rightline{\today} 
 
\begin{center} 
{\Large \bf NONCOMMUTATIVE ELECTRODYNAMICS AND ULTRA HIGH ENERGY GAMMA RAYS}\\ 
\vspace{2.cm} 
{\large Paolo Castorina$^{1,2}$, Dario Zappal\`a$^{2,1}$} \\ 
\vspace{.5cm}
{{\sl $^{1}$ Dipartimento di Fisica, Universit\`a di Catania} \\ 
{\sl Via S. Sofia 64, I-95123, Catania, Italy}\\ 
\vspace{.2cm} 
{\sl $^{2}$ INFN, Sezione di Catania,  Via S. Sofia 64, I-95123, Catania, Italy}} \\ 
 
\vspace{1.5 cm} 
{\sl e-mail: paolo.castorina@ct.infn.it; ~dario.zappala@ct.infn.it}\\ 
\vspace{2cm} 
 
\begin{abstract} 
Plane waves in  noncommutative classical electrodynamics  (NCED) have a peculiar dispersion relation. 
We investigate the  kinematical conditions on this deformed "mass shell" which come from ultra high energy  
gamma rays and  discuss noncommutative dynamical  effects on the gamma absorption by the infrared  
background and on the intrinsic spectrum. 
Finally  we note that in NCED there is a strong correlation between the modified dispersion relation 
and the presence of dynamical effects in electromagnetic phenomena such as in the case of  the  
synchrotron radiation.  
From this point of view, the   limits on the typical energy scale of the violation of Lorentz invariance  
obtained by deformed dispersion relations and by assuming undeformed dynamical effects should be  
taken with some caution. 
\end{abstract} 
\end{center} 
 
\vspace{2 cm} 
 
 
\end{titlepage}

\newpage 
Recently there has been a growing interest in theories where the speed of light is different from $c$ \cite{luce}. 
The main motivations of considering  this 'heresy' come from cosmology, quantum gravity and observation and  
indeed, ultra high energy cosmic (UHECR) and gamma (UHEGR) rays have been detected with energies which seem  
to be inconsistent with the standard GZK cutoff \cite{gzk}. 
 
It has been suggested that a possible explanation of the observed high energy cosmic rays could be  a modified dispersion 
relation  among energy, momentum and mass due to physical phenomena at Plank scale \cite{amelino0,amelino1,amelino2,liberati2,liberati3,steckern}  
or to explicit Lorentz invariant breaking terms in the lagrangian, originally proposed in \cite{jackli} and reconsidered more recently in 
\cite{coleman}.  
 
Also in noncommutative electrodynamics (NCED) \cite{jackiw}  
one has a modified dispersion relation for the electromagnetic waves and in this brief note  
we address the question whether it is compatible with the experimental data on  
UHEGR and we discuss some noncommutative dynamical effects in gamma rays absorption  
by the diffuse interstellar or intergalactic infrared radiation and in the $TeV$ photon  intrinsic production spectrum. 
 
 The simplest case of NCED  is described by  the action (the fermionic field is omitted): 
\begin{equation}  \label{othetamaxwell} 
\hat{I} = - \frac{1}{4} \int d^4 x \; [F^{\mu \nu} F_{\mu \nu} 
-\frac{1}{2} \theta^{\alpha \beta} F_{\alpha \beta} F^{\mu \nu} 
F_{\mu \nu} + 2 \theta^{\alpha \beta} F_{\alpha \mu} F_{\beta \nu} 
F^{\mu \nu}] \;, 
\end{equation} 
where the noncommutativity of space-time coordinates 
has been expressed in the canonical form \cite{wess}, by 
\begin{equation}  \label{due} 
x^\mu * x^\nu - x^\nu * x^\mu = i \theta^{\mu \nu} \;, 
\end{equation} 
the $*$-product is the standard  Moyal product (\cite{review}); 
$\hat{F}_{\mu \nu} = \partial_\mu \hat{A}_\nu - \partial_\nu 
\hat{A}_\mu - i [ \hat{A}_\mu , \hat{A}_\nu ]_* $ 
and 
$\hat{A}_\mu$ 
has been expressed in terms of a U(1) gauge field $A_\mu$ by 
 the $O(\theta)$ Seiberg-Witten \cite{sw} map 
\begin{equation}\label{tre} 
\hat{A}_{\mu}(A, \theta) = A_{\mu} - \frac{1}{2} \theta^{\alpha 
\beta}A_{\alpha}(\partial_{\beta}A_{\mu} + F_{\beta \mu})  \;. 
\end{equation} 
In \cite{jackiw} it was found that, in the presence of a 
background magnetic field $\vec{b}$, the $O(\theta)$ plane-wave 
classical 
solutions exist. The waves propagating transversely to $\vec{b}$ 
have a modified dispersion relation 
\begin{equation}\label{quattro} 
 \omega/c = k (1 - \vec{\theta}_T \cdot 
\vec{b}_T) 
\end{equation} 
(where $\vec{\theta} \equiv (\theta^{1}, \theta^{2}, 
\theta^{3})$, with $\theta^{i j} = \epsilon ^{i j k} \theta_{k}$, and 
$\theta^{0 i} = 0$) while the ones propagating along the direction 
of $\vec{b}$ still travel at the usual speed of light $c$. 
 
For completeness, let us remark that the quantum theory of noncommutative  
electrodynamics is still not completely understood.  
The perturbative calculations show a novel feature of the theory,  
called infrared-ultraviolet (IR-UV) connection\cite{review},  
that prevents to take the limit of the noncommutative parameter to zero and,  
moreover, the photon self energy introduces, at one loop order  
in perturbation theory, a tachyonic pole which goes to minus infinity. 
However, in our opinion the IR-UV connection 
could be for instance an effect induced by the  
application of standard  perturbative techniques or by  
a restricted theoretical framework. 
For example in \cite{chu} (see also \cite{ydri})  it is shown  
that there are (scalar) field theories where the IR-UV connection  
is absent and its presence depends on the projection on the  
noncommutative plane. 
Due the present status of the quantum formulation of the noncommutative theory,  
we assume the conservative point of view of working in the framework of the classical 
noncommutative theory which, for $\vec \theta\to 0$, reproduces the standard  
electrodynamics. 
 
The new dispersion relation in Eq. (\ref{quattro})  has many physical consequences which are 
however not easily observable. As a matter of fact, by using the bound of 
$\theta < 10^{-2} (TeV)^{-2}$ \cite{bound}, one would need a 
background magnetic field of the order of 1 Tesla over a distance 
of 1 parsec to appreciate the shift of the interference fringes 
due to the modified speed of noncommutative light. 
Mainly for this reason,  other classical and quantum phenomena have been 
recently proposed to improve the bound and to find new applications of 
noncommutativity \cite{loro}. 
 
It is then interesting to verify whether the deformed 
dispersion relation in Eq.(\ref{quattro}) is compatible with the astrophysical observations 
of UHEGR \cite{crab} \cite{extrag} and, more generally,  
what kind of phenomenological limits on the energy scale which characterizes the violation 
of the Lorentz invariance are obtained by a dynamical, and not purely kinematical, analysis 
of the NCED effects. 
 
Let us start by evaluating the effects of the dispersion relation 
in Eq.(\ref{quattro})  on the kinematical threshold for $\gamma  \rightarrow e^+ e^-$ and 
$\gamma + \gamma \rightarrow e^+ e^-$. 
 
In Eq.(\ref{quattro}) the  noncommutative contribution depends on the angle between the transverse components 
( with respect to $\vec k$) of the background magnetic field $\vec b$ and of the vector $\vec \theta$. 
At the first order in $\theta$ one has 
\be\label{cinque} 
{\omega}( 1 + \vec \theta _T \cdot \vec b_T) / c = k 
\ee 
and the ``mass shell'' relation becomes 
\be\label{sei} 
E_\gamma ^2 - k^2 = - 2 E_\gamma ^2 ( \vec \theta _T \cdot \vec b_T). 
\ee 
 where $ E_{\gamma} = \omega /c$ . 
 
In the case of NCED, with no modified dispersion relation for the electron, the decay 
$\gamma  \rightarrow e^+ e^-$ is kinematically permitted \cite{coleman}. On the other hand this 
decay is forbidden, for example, 
with the dispersion relation 
\be\label{sette} 
E ^2 - k^2 = -  E^3 / E_{QG} 
\ee 
which comes from quantum gravity effect with  typical energy scale $E_{QG}$ \cite{amelino0}. 
More general relations of the form in  Eq. (\ref{sette}) which include phenomenological parameters, have been considered in  
\cite{amelino1,liberati2}. 
 
By defining $p_\gamma$ the four momentum of the photon and $p_+$ and $p_-$ the four momenta of  $e^+$ and $e^-$, 
the kinematical condition for the decay is 
 
\be\label{otto} 
 - 2 E_\gamma ^2 ( \vec \theta _T \cdot \vec b_T) =( p_+ + p_-)^2 > 4 m_e^2 
\ee 
which requires  $( \vec \theta _T \cdot \vec b_T) < 0$. 
Since gamma rays with energy $\simeq 50~TeV$ have been observed 
\cite{crab} from the Crab Nebula this implies that 
 
\be\label{nove} 
(\vec \theta _T \cdot \vec b_T) > -  2 * 10^{-16} 
\ee 
The differences between the dispersion relation for NCED and Eq. (\ref{sette}) are that in our case 
the parameter  which modifies the relation can be negative, that is the speed of noncommutative light is less than $c$, 
and that there is a different dependence of the modification on the energy (quadratic vs. cubic). 
 
 Let us notice that our previous result is essentially 
the same kinematical limit obtained in \cite{coleman} 
 where however the physical mechanism is the difference in the maximum 
speed between the electron and the photon while in our case the mass shell relation for the electron is unchanged. 
Moreover in NCED the modification depends on an external parameter, the background magnetic field. Indeed, in the 
limits on $( \vec \theta _T \cdot \vec b_T)$ one has to specify if $\vec b$ is , for example, 
the galactic or extragalactic magnetic field. In discussing the limit from the observation of gamma rays from the Crab nebula 
$\vec b$ is the galactic magnetic field which is of order $ b_g \simeq 1 \mu G$. 
 
Let us now consider the limit which comes from $\gamma + \gamma_b \rightarrow e^+ e^-$ 
where $\gamma_b$ is a background low energy photon 
with momentum  $p_b= (E_b,\vec p_b)$. In this case the kinematical condition is 
 
 \be\label{dieci} 
- 2 E_\gamma ^2 ( \vec \theta _T \cdot \vec b_T) + 2 E_b E_\gamma 
- 2 E_b E_\gamma \cos \phi (1 + \vec \theta _T \cdot \vec b_T) 
 > 4 m_e^2 
\ee 
where the effect of the modified dispersion relation has been neglected 
for the low energy photon ( because numerically irrelevant) 
 and $\phi$ 
is the angle between  the momenta of the photons. 
Since the observation of extragalactic gamma rays up to energy $\simeq 20~TeV$ is in agreement with 
the absorption by the infrared diffuse extragalactic background 
\cite{mark}, let us consider the energy $E_b$ is in the range $1~meV - 200~meV$ 
to obtain the kinematical constraints on the noncommutative parameter.  
Then, for a central collision ($\cos \phi = -1$), the condition is 
 
 \be\label{undici} 
-  ( \vec \theta _T \cdot \vec b_T) > + 2 (m_e / E_\gamma)^2 - 2 (E_b / E_\gamma ) 
\ee 
 which for $E_b =1~meV$ gives a negative value 
 \be\label{dodici} 
( \vec \theta _T \cdot \vec b_T) < - 1.15  * 10^{-15} 
\ee 
while for $E_b =200~meV$  gives 
\be\label{tredici} 
( \vec \theta _T \cdot \vec b_T) < 1.9  * 10^{-14} 
\ee 
 
In combining the kinematical conditions in Eqs. (\ref{nove}), (\ref{dodici}) and (\ref{tredici})  
we have to take into account that  
the extragalactic background magnetic field $b_{eg}$ is about $10^{-3} ~b_{g}$. 
Then, for a far  infrared extragalactic absorption ( $E_b = 1~ meV$)  the conditions, Eq. (\ref{nove}) and Eq. (\ref{dodici}), cannot be simultaneously 
satisfied. Indeed, the combined constraints limit the range of $E_b$ for  absorption of UHEGR.  If one evaluates 
Eq. (\ref{undici})  with the value in the right hand side of Eq. (\ref{nove}), 
 it can be easily seen that the production $\gamma + \gamma_b \rightarrow e^+ e^-$ is possible only for energy of the background photon 
$E_b > O(10~meV)$. This value of  $E_b$ 
is close to that one which maximizes the standard electrodynamics pair production absorption cross section \cite{stecker2}. 
 
In the second case ( $E_b = 200~meV$) the conditions are 
 \be\label{quattordici} 
 -2  * 10^{-16} < \vec \theta _T \cdot \vec b_{gT} < 1.9 * 10^{-11} 
\ee

 If we consider the  limit  $|\theta | < (10~TeV)^{-2}$ and the magnitude of the galactic 
or extragalactic 
 magnetic field, the previous conditions are satisfied by many order of magnitudes: 
\be\label{quindici} 
| \vec \theta _T \cdot \vec b_{gT} | < |\theta _T | | \vec b_{gT}|  \simeq 10^{-30} 
 \ee 
 
Then the dispersion relation coming from NCED is largely consistent with the present data on UHEGR. 
The previous results could be an effect of the simplified version of the theory here considered. Indeed, in the complete  
version of NCED also a  modified  dispersion relation for electrons and positrons could possibly be present.  
However we do not believe that this effects would dramatically change 
the previous conclusions and moreover the problem of mass generation in noncommutative field theory  is still open \cite{noi2} 
as well as  the possibility to have a full consistent quantum  field theory \cite{quantum}. 
 
In addition to the kinematical constraints discussed above,  there are many other noncommutative dynamical effects for example in the  
absorption cross section and in the source gamma ray spectrum. 
In fact  the standard calculations to fit the observed gamma ray data \cite{stecker2} require the evaluation of the optical depth 
for attenuation between the source and the Earth due to  the $\gamma + \gamma_b \rightarrow e^+ e^-$ cross section 
and the most widely investigated model for the production of $TeV$ photons 
involves the injection of relativistic electrons  via the synchrotron self-Compton mechanism \cite{ssc}. 
  
While from  the perturbative quantum noncommutative calculations, where  
the modifications in the dispersion relations are neglected,  a very small effect to the $\gamma + \gamma_b \rightarrow e^+ e^-$  
process is predicted \cite{radcor} in the energy range  
$E_b\sim 100~meV$, $E_\gamma\sim 20~TeV$ with  $|\theta|<(10~TeV)^{-2}$,   
it has been shown in \cite{noi} that a consequence of the modified dispersion relation, as in Eq. (\ref{quattro}), is a large correction to the {\it classical}  
synchrotron spectrum of  a  relativistic particle in a strong magnetic field.  
As an example,  the correction to the synchrotron spectrum for a relativistic electron  
to $O(\theta) $, for $|\theta | < (10~TeV)^{-2}$ and for $\omega_0<<\omega<<\omega_0 \gamma^3 $ turns out \cite{noi} 
 
\begin{equation}\label{sedici} 
X=\frac{d I (\omega) / d \Omega}{d I(\omega) / d 
\Omega|_{\theta = 0}}  
< 1 + \left ( \frac{\omega_0}{\omega} \right )^{2/3} n \times 10^{-21} \times 
(E_{electron} (MeV)/(MeV))^4 \;, 
\end{equation} 
where  $\omega_0$ is the synchrotron frequency, ${\omega}$ is the radiation frequency, $\gamma$ is the Lorentz factor, $E_{electron}$  
is the energy of the electron and $n$ is 
the value of the  magnetic field, which accelerates the electron, expressed in Tesla.  
If a 20 $TeV$ photon is produced by synchrotron radiation, the correction is at least 
\begin{equation}\label{diciasette} 
X < 1 + 1.6 \left ( \frac{\omega_0}{\omega} \right )^{2/3} n \times 10^{8}, 
\end{equation} 
The absolute value of the correction depends on the  frequencies and on the magnetic field one is considering, 
but this result is a signature that the modification of the dispersion relation is strongly correlated with  dynamical effects. 
From this point of view, the   limits on the typical 
energy scale of the violation of Lorentz invariance obtained by deformed dispersion relations 
and by assuming undeformed dynamical effects should be taken with some caution. 
 
In conclusion our analysis shows that the kinematical constraints form the UHEGR are too weak to give meaningful 
indications on the noncommutativity parameter $\theta$. Conversely, we expect that the NCED could play a  
relevant role in the dynamical processes involved in the intrinsic production spectrum of the UHEGR,  
as the sychrotron self-Compton mechanism, thus providing a possible mean to put a tight bound on $\theta$.

\vspace{0.6 cm} 
\leftline{\bf Acknowledgements} 
We are grateful to G. Nardulli for many indications and suggestions on the subject and to A. Iorio for stimulating discussions.

\end{document}